\documentclass{aastex61}

\received{\today}
\revised{}
\accepted{}
\submitjournal{ApJ}

\usepackage{graphicx}
\usepackage{dcolumn}
\usepackage{bm}
\usepackage{xspace}
\usepackage{soul}
\usepackage{hyperref}
\usepackage{amsmath}

\def\liv{\ifmmode {\mathrm{LIV}}\else{\scshape LIV}\fi\xspace}
\def\li{\ifmmode {\mathrm{LI}}\else{\scshape LI}\fi\xspace}

\shorttitle{\liv\ limits from UHECR}
\shortauthors{Lang, Mart\'inez-Huerta and de Souza}

\begin{document}

\title{Limits on the Lorentz Invariance Violation from UHECR astrophysics.}

\correspondingauthor{Rodrigo Guedes Lang}
\email{rodrigo.lang@usp.br}

\author{Rodrigo Guedes Lang}
\affiliation{Instituto de F\'isica de S\~ao Carlos, Universidade de S\~ao Paulo, Av. Trabalhador S\~ao-carlense 400, S\~ao Carlos, Brasil.}

\author{Humberto Mart\'inez-Huerta}
\affiliation{Instituto de F\'isica de S\~ao Carlos, Universidade de S\~ao Paulo, Av. Trabalhador S\~ao-carlense 400, S\~ao Carlos, Brasil.}
\affiliation{Departamento de F\'isica, Centro de Investigaci\'on y de Estudios Avanzados del I.P.N., Apartado Postal 14-740, 07000, Ciudad de M\'exico, M\'exico.}

\author{Vitor de Souza}
\affiliation{Instituto de F\'isica de S\~ao Carlos, Universidade de S\~ao Paulo, Av. Trabalhador S\~ao-carlense 400, S\~ao Carlos, Brasil.}

\begin{abstract}
In this paper, Lorentz Invariance Violation (\liv) is introduced in the calculations of photon propagation in the Universe. \liv is considered in the photon sector and the mean free path of the $\gamma \gamma \rightarrow e^{+} e^{-}$ interaction is calculated. The corresponding photon horizon including \liv effects is used to predict major changes in the propagation of photons with energy above $10^{18}$ eV. The flux of GZK photons on Earth considering \liv is calculated for several source models of ultra-high energy cosmic ray (UHECR). The predicted flux of GZK gamma-rays is compared to the new upper limits on the photon flux obtained by the Pierre Auger Observatory in order to impose upper limits on the \liv coefficients of order $n =$ 0, 1 and 2.  The  limits on the \liv coefficients derived here are more realistic than previous works and in some cases more restrictive. The analysis resulted in \liv upper limits in the photon sector of $\delta_{\gamma,0}^{limit} \sim -10^{-20}$, $\delta_{\gamma,1}^{limit} \sim -10^{-38} \; \mathrm{eV^{-1}}$ and $\delta_{\gamma,2}^{limit} \sim -10^{-56} \; \mathrm{eV^{-2}}$ in the astrophysical scenario which best describes UHECR data.
\end{abstract}

\keywords{cosmic rays - Lorentz invariance violation}

\section{Introduction}

Astroparticle physics has recently reached the status of precision science due to: a) the construction of new observatories operating innovative technologies, b) the detection of large numbers of events and sources and c) the development of clever theoretical interpretations of the data. Two observational windows have produced very important results in the last decade. The ultra-high energy cosmic rays (E $>$ EeV) studied by the Pierre Auger and the Telescope Array Observatories (\cite{ThePierreAugerCollaboration2015172,Tinyakov201429}) improved our knowledge of the most extreme phenomena known in Nature. The GeV-TeV gamma-ray experiments FERMI/LAT (\cite{bib:fermi:lat}), H.E.S.S. (\cite{bib:hess}), MAGIC (\cite{bib:magic}) and VERITAS (\cite{bib:veritas}) gave a new perspective on gamma-ray production and propagation in the Universe. The operation of the current instruments and the construction of future ones (\cite{bib:cta,bib:jem:euso,1674-1137-34-2-018}) guarantee the production of even more precise information in the decades to come.

Lorentz Invariance (\li) is one of the pillars of modern physics and it has been tested in several experimental approaches(\cite{bib:liv:tests}). Astroparticle physics has been proposed as an appropriate test environment for possible Lorentz Invariance Violation (\liv) given the large energy of the particles, the large propagation distances, the accumulation of small interaction effects and recently the precision of the measurements (\cite{bib:liv:tests:astropart,Stecker:2004, Stecker:2009, AmelinoEllis:1998, Jacobson:2002, Gunter:2007, Gunter:2008,XU201672,1674-1137-40-4-045102,ELLIS201350,ALBERT2008253,Ellis2006402,ELLIS2008158,Farbairn:2014,Biteau:2015,Tavecchio:2015,Rubtosov:2017}).

Effective field theories with some Lorentz violation can derive in measurable effects in the data taking by astroparticle physics experiments, nonetheless, in this paper \liv is introduced in the astroparticle physics phenomenology through the polynomial correction of the dispersion relation in the photon sector, and is focused on the gamma-ray propagation and  pair production effects with \liv. Other phenomena like vacuum birefringence, photon decay, vacuum Cherenkov radiation, photon splitting, synchrotron radiation and helicity decay have also been used to set limits on LIV effects on the photon sector but are beyond the scope of this paper, for a review see~\cite{bib:liv:tests:astropart, Bluhm:2014, Rubtosov:2017}.

Lorentz invariant  gamma-ray propagation in the intergalactic photon background was studied previously in detail by~\cite{bib:angelis}, a similar approach is followed in section~\ref{sec:horizon}, but \liv is allowed in the interaction of high energy photons with the background light and their consequences are studied.  The process $\gamma \gamma \rightarrow e^{+} e^{-}$ is the only one considered to violate Lorentz invariance, and as a similar approach used in~\cite{Gunter:2007}, such LIV correction can lead to a correction of the \li energy threshold of the production process. The latter phenomena modifies the mean free path of the interaction and therefore the survival probability of a photon propagating through the background light, which depends on the \liv coefficients. This dependence is calculated in section~\ref{sec:horizon} and the mean free path and the photon horizon are shown for several \liv coefficients and different orders of the \liv expansion in the photon energy dispersion relation.

In section~\ref{sec:gzk:photons}, the mean free path of the photo-production process considering \liv is implemented in a Monte Carlo propagation code in order to calculate  the effect of the derived \liv in the flux of ultra-high energy photons arriving on Earth due to the GZK effect~(\cite{bib:g,bib:zk}) and considering several models for the sources of cosmic rays.  Section~\ref{sec:gzk:photons} quantifies the influence of the astrophysical models concerning mass composition, energy spectra shape and source distribution. These dependencies have been largely neglected in previous studies and it is shown here that they influence the GZK photon flux by as much as four orders of magnitude.

In section~\ref{sec:limits}, the propagated GZK photon flux for each model is compared to recent upper limits on the flux of photons obtained by the Pierre Auger Observatory. For some astrophysical models, the Auger data is used to set restrictive limits on the \liv coefficients. The astrophysical model used to describe the primary cosmic ray flux has a very large influence on the flux of GZK photons and therefore on the \liv limits imposed.  Finally, in section~\ref{sec:conclusions} the conclusions are presented.


\section{Photon horizon including \liv effects}
\label{sec:horizon}

One of the most commonly used mechanisms to introduce \liv in particle physics phenomenology is based on the polynomial correction in the dispersion relation of a free propagating particle, mainly motivated by an extra term in the Lagrangian density that explicitly breaks Lorentz symmetry, see for instance references~\cite{AmelinoEllis:1998,Glashow:1999,Ahluwalia:1999,Amelino:2001, Jacobson:2002,Gunter:2007,Gunter:2008,Liberati:2008,Liberati:2011,1475-7516-2008-01-031,refId1}. In these models, the corrected expression for the dispersion relation is given by the following equation:
\begin{equation}
E_{a}^2 - p_{a}^2 = m_{a}^2 + \delta_{a,n} E_{a}^{n+2},
\label{eq:dispersion}
\end{equation}
where $a$ denotes the particle with mass $m_a$ and four-momenta $(E_{a},p_{a})$. For simplicity, natural units are used in this work. The \liv coefficient, $\delta_{a,n}$, parametrizes the particle dependent \liv correction, where $n$ expresses the correction order, which can be derived from the series expansion or from a particular model for such order, see for instance the case of $n=0$ (\cite{GLASHOW:97,Glashow:1999,Example_n0}), $n=1$ (\cite{Example_n1}) or for a generic $n$ (\cite{bib:fermigrb}). The \liv parameter of order $n$, $\delta_n$, is frequently considered to be inversely proportional to some \liv energy scale $E_{LIV}^{(n)}$. Different techniques have been implemented in the search of LIV signatures in astroparticle physics and some of them have been used to derive strong constraints to the \liv energy scale (\cite{AmelinoEllis:1998,Liberati:2008,bib:hesspks,bib:fermigrb,VERITAS-LIV2,PhysRevD.79.083015,Otte:2011,Schreck:2014,Biteau:2015,HMartinez:2016,Rubtosov:2017}).

The threshold analysis of the pair production process, considering the \liv corrections from equation 1 on the photon sector is discussed in appendix~\ref{sec:liv:model} and leads to corrections of the LI energy threshold of the process. In the following, $\epsilon^{LIV}_{th}$ stands for the minimum energy of the  cosmic background (CB) photon in the pair production process with LIV.  The latter effect can lead to changes in the optical depth, $\tau_{\gamma} (E_\gamma,z)$, that quantifies how opaque to photons the Universe is. The survival probability, i.e., the probability that a photon, $\gamma$, emitted with a given energy, $E_\gamma$, and at a given redshift, $z$, reaches Earth without interacting with the background, is given by:
\begin{equation}
\centering
P_{\gamma \rightarrow \gamma} (E_\gamma,z) = e^{-\tau_{\gamma} (E_\gamma,z)}.
\end{equation}

The photon horizon is the distance ($z_h$) for which $\tau_{\gamma} (E_\gamma,z_h) = 1$. $z_h$ defines, as a function of the energy of the photon, the redshift at which a emitted photon will have probability $P_{\gamma \rightarrow \gamma} = 1/e$ of reaching Earth. The evaluation of the photon horizon is of extreme importance because it summarizes the visible Universe as a function of the energy of the emitted photon. In this section, the photon horizon is calculated including \liv effects. The argument presented in reference~\cite{bib:angelis} is followed here.

In the intergalactic medium, the $\gamma\gamma_{CB}\rightarrow e^+ e^-$ interaction is the main contribution to determine the photon horizon. In the approximation where cosmological effects are negligible, the mean free path, $\lambda (E_\gamma)$, of this interaction is given by:
\begin{equation}
\centering
\lambda (E_\gamma) = \frac{c z}{H_0 \tau_{\gamma} (E_\gamma,z)},
\end{equation}
where $H_{0} = 70$ km s$^{-1}$ Mpc$^{-1}$ is the Hubble constant and $c$ is the speed of light in vacuum. The optical depth is obtained by:
\begin{equation}
\label{eq:cosm}
\centering
\begin{split}
\tau_{\gamma} (E_{\gamma},z) = \int_{0}^{z} dz \frac{c}{H_{0} (1+z) \sqrt{\Omega_{\Lambda} + \Omega_{M} (1+z)^{3}}} \times
\int_{-1}^{1} d(\cos\theta) \frac{1-\cos{\theta}}{2} \int_{\epsilon_{th}^{\liv}}^{\infty} d\epsilon n_{\gamma} (\epsilon,z) \sigma(E_\gamma,\epsilon,z),
\end{split}
\end{equation}
where $\theta$ is the angle between the direction of propagation of both photons $\theta = [-\pi,+\pi]$, $\Omega_{\Lambda} = 0.7$ is the dark energy density, $\Omega_{M} = 0.3$ is the matter density, $\sigma$ is the cross-section of the interaction and $\epsilon_{th}^{\liv}$ is the threshold energy of the interaction as given by equation~\ref{eq:e:cons:th}.

$n_{\gamma_{CB}}$ is the background photon density. The dominant backgrounds are the Extra-galactic Background Light (EBL) for $E_\gamma < 10^{14.5}$ eV, the Cosmic Background Microwave Radiation (CMB) for $10^{14.5} \ \mathrm{eV} < E_\gamma < 10^{19}$ eV and the Radio Background (RB) for $E_\gamma > 10^{19}$ eV. In the calculations presented here, the Gilmore model (\cite{bib:gilmore}) was used for the EBL. Since \liv effects in the photon horizon are expected only at the highest energies ($E_\gamma > 10^{16}$ eV) using different models of EBL would not change the results. For the RB, the data from \cite{bib:gervasi} with a cutoff at 1 MHz were used. Different cutoffs in the RB data lead to different photon horizons as shown in reference~\cite{bib:angelis}. Since no new effect shows up in the \liv calculation due to the RB cutoff, only the 1 MHz cutoff will be presented.

It is usual for studies such as the one presented here, in which the threshold of an interaction is shifted causing a modification of the mean free path, to neglect direct effects in the cross section, $\sigma$, when solving equation~\ref{eq:cosm}. However an implicit change of the cross section is taken into account given its dependence on the energy threshold $\epsilon_{th}^{\liv}$ (\cite{bib:breit:wheeler}).

%
%
%
%
%

\begin{figure}[!ht]
	\centering
	\includegraphics[width=.48\textwidth]{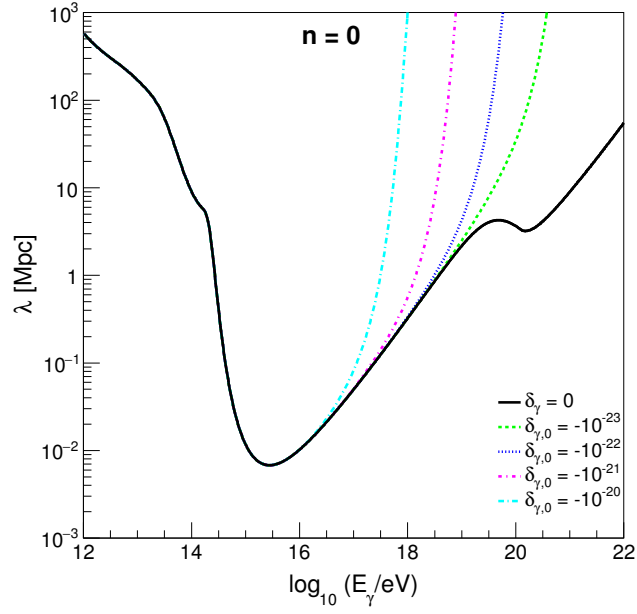}
	\caption{Mean free path ($\lambda$) for $\gamma \gamma_{CB} \rightarrow e^{+} e^{-}$ as a function of the energy of the photon ($E_\gamma$) shown for several \liv coefficients for $n=0$. The Gilmore model (\cite{bib:gilmore}) for EBL and  Gervasi et al. (\cite{bib:gervasi}) model for the RB with a cutoff at 1 MHz were used. The black continuous line is the \li scenario. The colored lines represent different values for the \liv coefficients. The colored lines coincide with the black line for $\log(E_\gamma/eV) < 15$.}
	\label{fig:LIVpathn0}
\end{figure}

\begin{figure}[!ht]
    \centering
    \includegraphics[width=.48\textwidth]{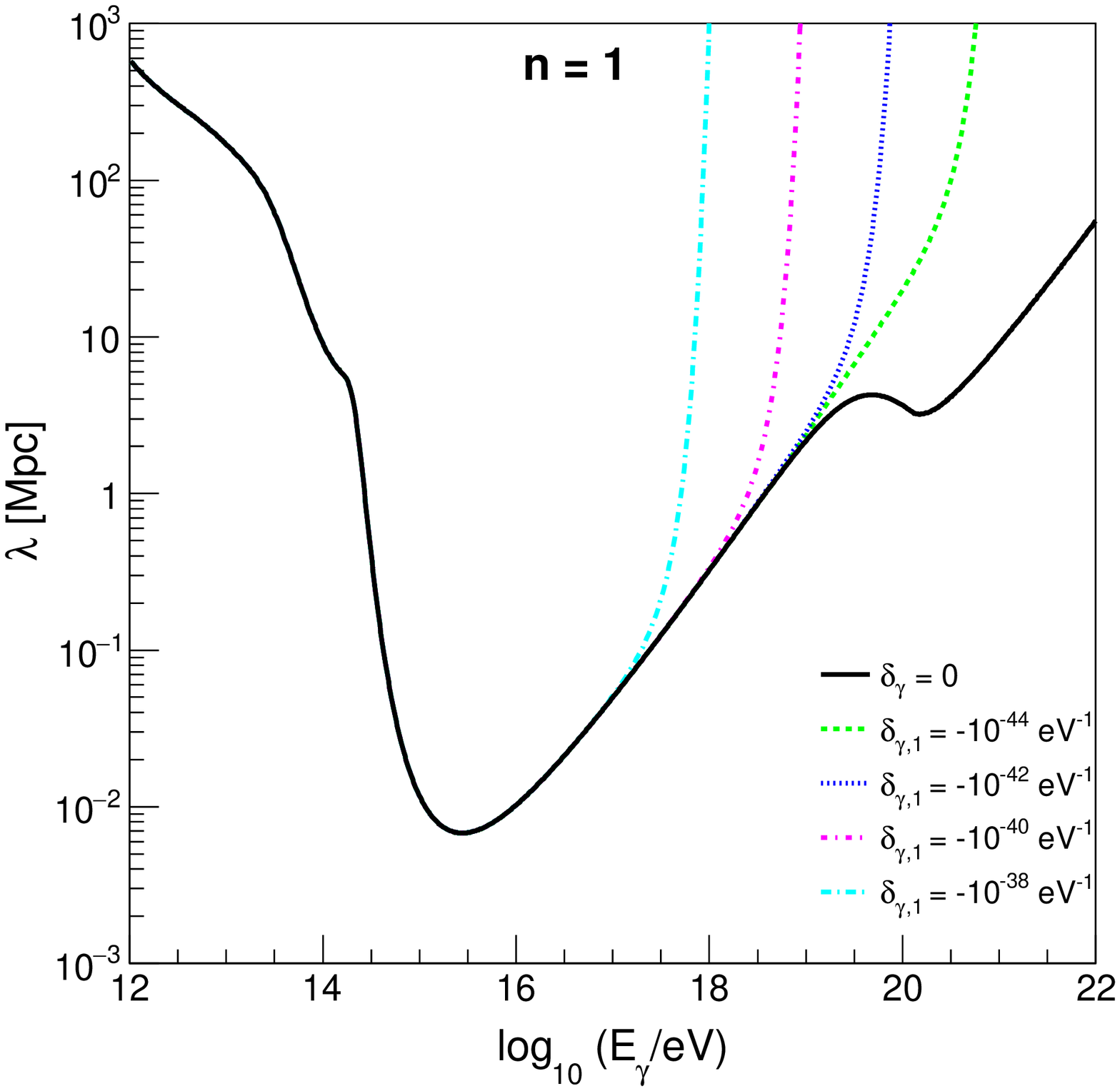}
	\caption{Mean free path ($\lambda$) for $\gamma \gamma_{CB} \rightarrow e^{+} e^{-}$ as a function of the energy of the photon ($E_\gamma$) shown for several \liv coefficients for $n=1$. The Gilmore model (\cite{bib:gilmore}) for EBL and  Gervasi et al. (\cite{bib:gervasi}) model for the RB with a cutoff at 1 MHz were used. The black continuous line is the \li scenario. The colored lines represent different values for the \liv coefficients. The colored lines coincide with the black line for $\log(E_\gamma/eV) < 15$.}
	\label{fig:LIVpathn1}
\end{figure}

\begin{figure}[!ht]
	\centering
	\includegraphics[width=.48\textwidth]{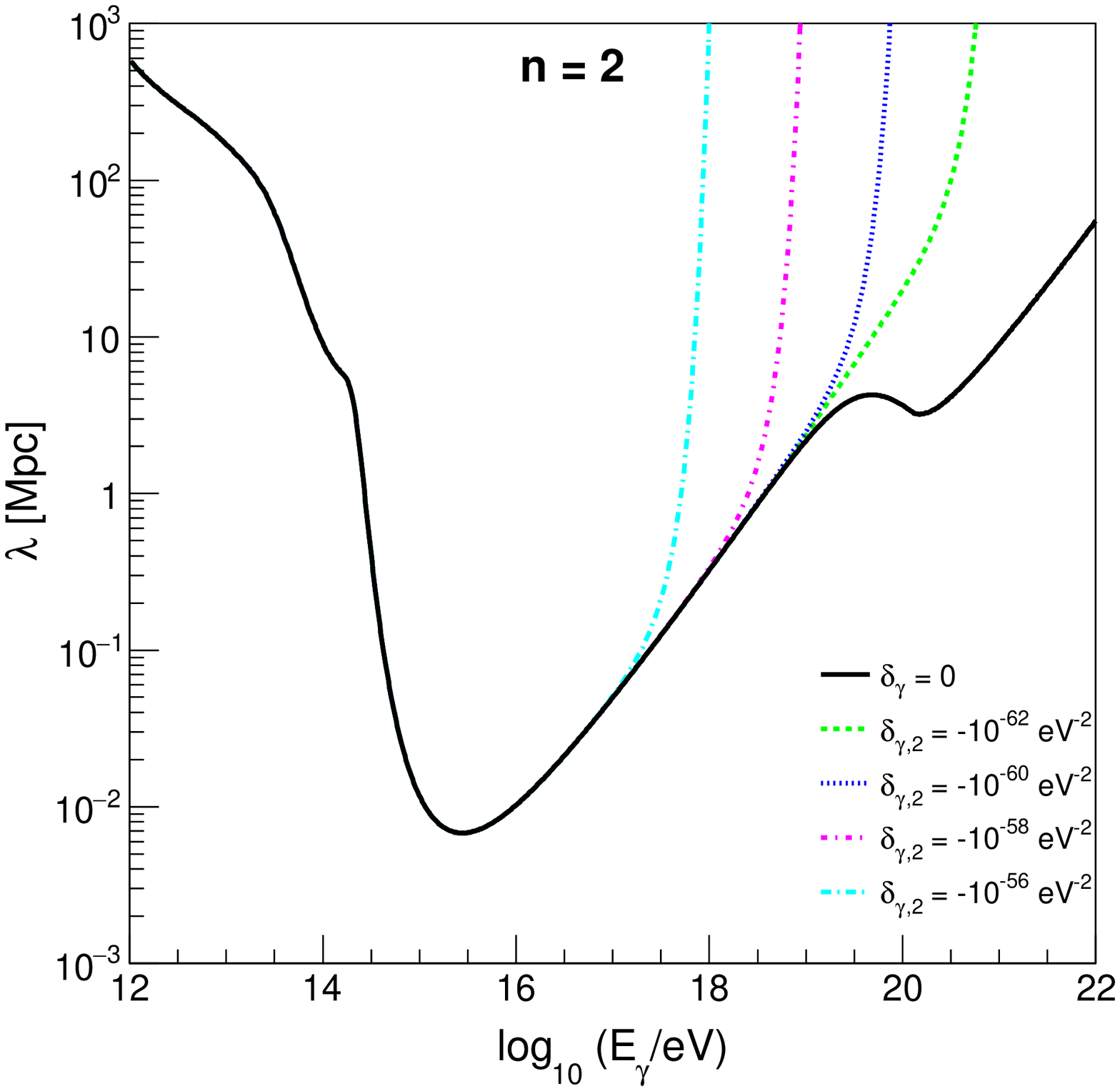}
	\caption{Mean free path ($\lambda$) for $\gamma \gamma_{CB} \rightarrow e^{+} e^{-}$ as a function of the energy of the photon ($E_\gamma$) shown for several \liv coefficients for $n=2$. The Gilmore model (\cite{bib:gilmore}) for EBL and  Gervasi et al. (\cite{bib:gervasi}) model for the RB with a cutoff at 1 MHz were used. The black continuous line is the \li scenario. The colored lines represent different values for the \liv coefficients. The colored lines coincide with the black line for $\log(E_\gamma/eV) < 15$.}
    \label{fig:LIVpathn2}
\end{figure}

Figures~\ref{fig:LIVpathn0},~\ref{fig:LIVpathn1} and~\ref{fig:LIVpathn2} show the mean free path for $\gamma \gamma_{CB} \rightarrow e^{+} e^{-}$ as a function of the energy of the photon, $E_\gamma$, for several \liv coefficients with $n=0$, $n=1$ and $n=2$, respectively. The main effect is an increase in the mean free path that becomes stronger the larger the photon energy, $E_\gamma$, and the \liv coefficient are. Consequently, fewer interactions happen and the photon, $\gamma$, will have a higher probability of traveling farther than it would have in a \li scenario. Similar effects due to \liv are seen for $n=0$, $n=1$ and $n=2$. The \liv coefficients are treated as free parameters, therefore there is no way to compare the importance of the effect between the orders $n=0$, $n=1$ and $n=2$, each order must be limited independently. Note that $\delta_{\gamma,n}$ units depend on $n$.

\begin{figure}[!ht]
	\centering
	\includegraphics[width=.48\textwidth]{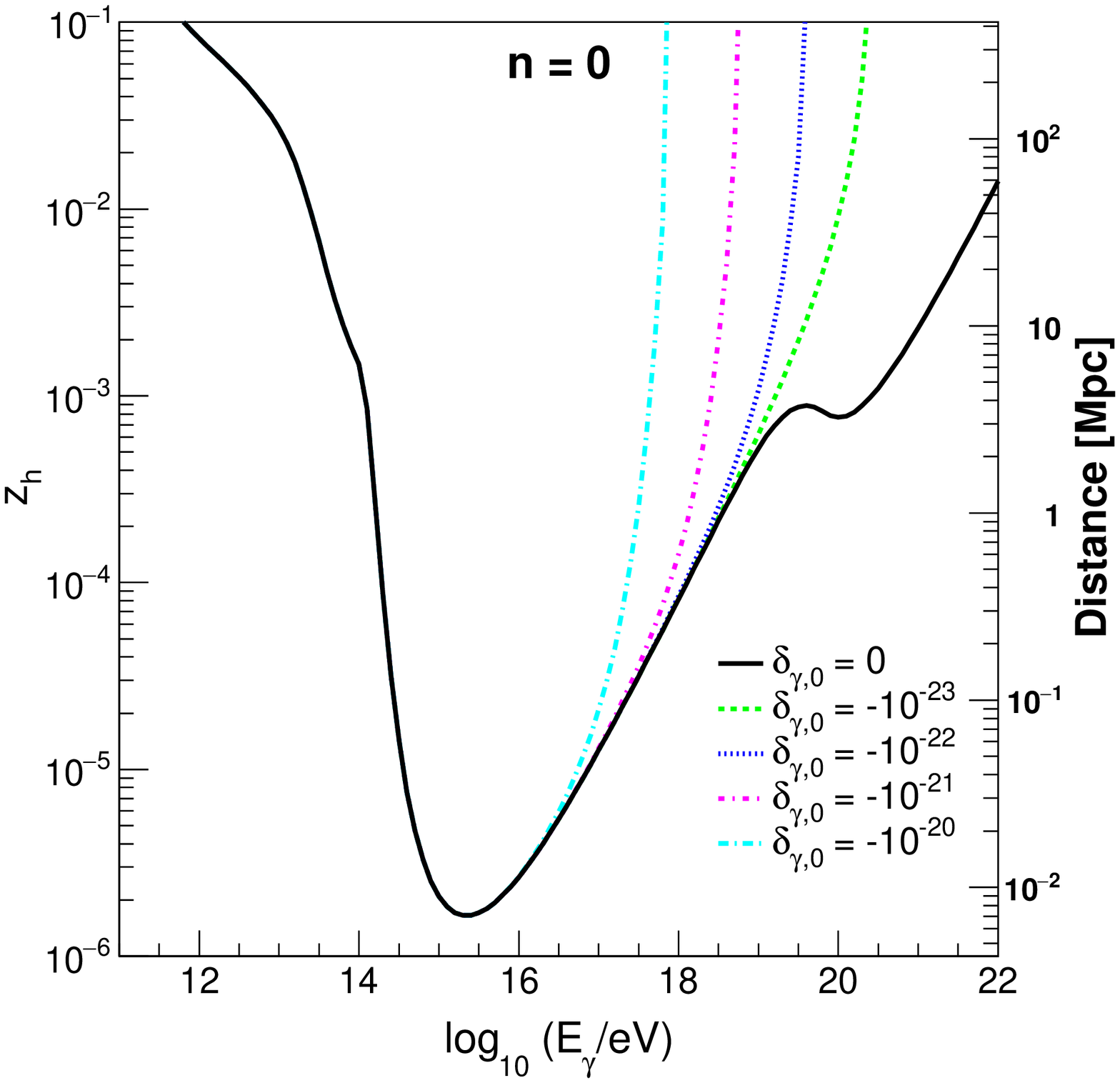}
	\caption{Photon horizon ($z_h$) as a function of the photon energy ($E_\gamma$) for different \liv coefficients with $n=0$. The right axis shows the equivalent distance obtained using the same assumptions used in equation \ref{eq:cosm}. The Gilmore model (\cite{bib:gilmore}) for EBL and  Gervasi et al. (\cite{bib:gervasi}) model for the RB with a cutoff at 1 MHz were used. The black continuous line represents the \li scenario. The colored lines represent different values for the \liv coefficients. The colored lines coincide with the black line for $\log(E_\gamma/eV) < 15$.}
	\label{fig:LIVhorizon}
\end{figure}

The \liv effect becomes more tangible in figure~\ref{fig:LIVhorizon} in which the photon horizon ($z_h$) is shown as a function of $E_\gamma$ for $n = 0$. For energies above $E_\gamma > 10^{16.5}$ eV and the given \liv values, the photon horizon increases when \liv is taken into account, increasing the probability that a distant source emitting high energy photons produces a detectable flux at Earth. Similar results are found for $n = 1$ and $n = 2$.

\section{Flux of GZK photons including \liv effects}
\label{sec:gzk:photons}

Even though the effects of \liv on the propagation of high energy photons are strong, they cannot be directly measured and, therefore, used to probe \liv models. In order to do that, in this section, the flux of GZK photons on Earth considering \liv is obtained and compared to the upper limits on the photon flux from the Pierre Auger Observatory (\cite{UpperLimitsAuger,UpperLimitsSD}).

UHECRs interact with the photon background producing pions (photo-pion production). Pions decay shortly after production generating EeV photons among other particles. The effect of this interaction chain suppresses the primary UHECR flux and generates a secondary flux of photons (\cite{Gelmini2007390}). The effect was named GZK after the authors of the original papers (\cite{bib:g,bib:zk}). The EeV photons (GZK photons) also interact with the background photons as described in the previous sections.

In order to consider \liv in the GZK photon calculation the CRPropa3/Eleca (\cite{bib:crpropa:3,Settimo}) codes were modified. The mean free paths calculated in section~\ref{sec:horizon} were implemented in these codes and the propagation of the particles was simulated. The resulting flux of GZK photons is, however, extremely dependent on the assumptions about the sources of cosmic rays, such as the injected energy spectra, mass composition, and the distribution of sources in the Universe. Therefore, four different models for the injected spectra of cosmic rays at the sources and five different models for the evolution of sources with redshift are considered in the calculations presented below.

\subsection{Models of UHECR sources}

No source of UHECR was ever identified and correlations studies with types of source are not conclusive. Several source types and mechanisms of particle production have been proposed. The amount of GZK photons produced in the propagation of the particles depends significantly on the source model used. In this paper, four UHECRs source models are used to calculate the corresponding GZK photons. The models are used as illustration of the differences in the production of GZK photons, an analysis of the validity of the models and its compatibility with experimental data is beyond the scope of this paper. However, it is important to note that strong constrains to the source models can be set by new measurements (\cite{bib:auger:combined:fit}). The models used here are labeled as:

\begin{itemize}
\item{\textbf{$\bm{C_{1}}$:} Aloisio, Berezinsky \& Blasi (2014) (\cite{bib:C1});}
\item{\textbf{$\bm{C_{2}}$:} Unger, Farrar \& Anchordoqui (2015) - Fiducial model (\cite{bib:C2});}
\item{\textbf{$\bm{C_{3}}$:} Unger, Farrar \& Anchordoqui (2015) (\cite{bib:C2}) with the abundance of galactic nuclei from (\cite{bib:C3});}
\item{\textbf{$\bm{C_{4}}$:} Berezinsky, Gazizov \& Grigorieva (2007) - Dip model (\cite{bib:C4}).}
\end{itemize}

All four models propose the energy spectrum at the source to be a power law distribution on the energy with a rigidity cutoff:
\begin{equation}
\centering
\frac{dN}{dE_{s}} = \begin{cases}
E_{s}^{-\Gamma} \ \ \ \ \ \ \ \ \ \ \ \ \ \ \ \text{, for } R_{s} < R_{cut} \\
E_{s}^{-\Gamma} e^{1 - R_{s}/R_{cut}} \ \text{, for } R_{s} \ge R_{cut}
\end{cases} ,
\end{equation}
where the spectral index, $\Gamma$, and the rigidity cutoff, $R_{cut}$, are parameters given by each model. Five different species of nuclei (H, He, N, Si and Fe) are considered in these models and their fraction ($f$H, $f$He, $f$N, $f$Si and $f$Fe) are given in Table~\ref{table:models}.

\begin{table}[!ht]
  \centering
  \begin{tabular}{ c | c c | c c c c c }
    \hline
    \hline
    Model & $\Gamma$ & $log_{10} (R_{cut}/V)$ & $f$H & $f$He & $f$N & $f$Si & $f$Fe \\
    \hline
    $C_{1}$ & 1 & 18.699 & 0.7692 & 0.1538 & 0.0461 & 0.0231 & 0.00759 \\
    $C_{2}$ & 1 & 18.5 & 0 & 0 & 0 & 1 & 0 \\
    $C_{3}$ & 1.25 & 18.5 & 0.365 & 0.309 & 0.121 & 0.1066 & 0.098 \\
    $C_{4}$ & 2.7 & $\infty$ & 1 & 0 & 0 & 0 & 0 \\
\hline
\hline
  \end{tabular}
  \caption{Parameters of the four source models used in this paper. $\Gamma$ is the spectral index, $R_{cut}$ is the rigidity cutoff and $f$H, $f$He, $f$N, $f$Si and $f$Fe are the fractions of each nuclei.}
  \label{table:models}
\end{table}

The composition of UHECR has a strong influence on the generated flux GZK photons and, therefore, on the possibility to set limits on \liv effects. The models chosen in this study ranges from very light ($C_{4}$) to very heavy ($C_{2}$) passing by intermediate compositions $C_1$ and $C_3$. Heavier compositions produces less GZK photons and therefore as less prone to reveal \liv effects.

\begin{figure}[!h]
	\centering
	\includegraphics[width=.48\textwidth]{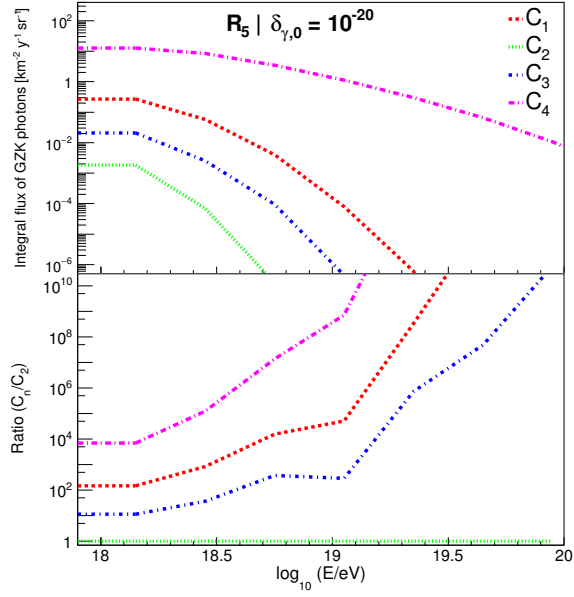}
	\caption{Integral flux of GZK photons as a function of the photon energy for each source model. Each line represents a different model $C_{n}$. All cases are for the source evolution model $R_{5}$ and \liv coefficient $\delta_{\gamma,0} = 10^{-20}$. The top panel shows the integral flux, while the bottom panel show the ratio to the one that produces less photons, $C_{2}$.}
	\label{fig:source:models}
\end{figure}

Figure~\ref{fig:source:models} shows the dependence of the GZK photon flux on the source model used. The integral of the GZK photon fluxes for \liv case of $\delta_{\gamma,0} = 10^{-20}$ are shown as a function of energy. The use of different \liv coefficients results in a shift up an down in the integral flux for each source model, having negligible changes in each ratio. The dependence on the model is of several orders of magnitude and should be considered in studies trying to impose limits on \liv coefficients. The capability to restrict \liv effects is proportional to the GZK photon flux generated in each model assumption.

\subsection{Models of source distribution}

Figure~\ref{fig:LIVhorizon} shows how the photon horizon increases significantly when \liv is considered. Therefore the source distribution in the Universe is an important input in GZK photon calculations usually neglected in previous studies. Five different models of source evolution($R_{n}$) are considered here:
\begin{itemize}
\item{\textbf{$\bm{R_{1}}$:} Sources are uniformly distributed in a comoving volume;}
\item{\textbf{$\bm{R_{2}}$:} Sources follow the star formation distribution given in reference~\cite{bib:R2}. The evolution is proportional to $(1+z)^{3.4}$ for $z < 1$, to $(1+z)^{-0.26}$ for $1 \le z < 4$ and to $(1+z)^{-7.8}$ for $z \ge 4$;}
\item{\textbf{$\bm{R_{3}}$:} Sources follow the star formation distribution given in reference~\cite{bib:R3}. The evolution is proportional to $(1+z)^{3.4}$ for $z < 1$, to $(1+z)^{-0.3}$ for $1 \le z < 4$ and to $(1+z)^{-3.5}$ for $z \ge 4$;}
\item{\textbf{$\bm{R_{4}}$:} Sources follow the GRB rate evolution from reference~\cite{bib:R4}. The evolution is proportional to $(1+8z)/[1+(z/3)^{1.3}]$;}
\item{\textbf{$\bm{R_{5}}$:} Sources follow the GRB rate evolution from reference~\cite{bib:R4}. The evolution is proportional to $(1+11z)/[1+(z/3)^{0.5}]$.}
\end{itemize}

\begin{figure}[!h]
	\centering
	\includegraphics[width=.48\textwidth]{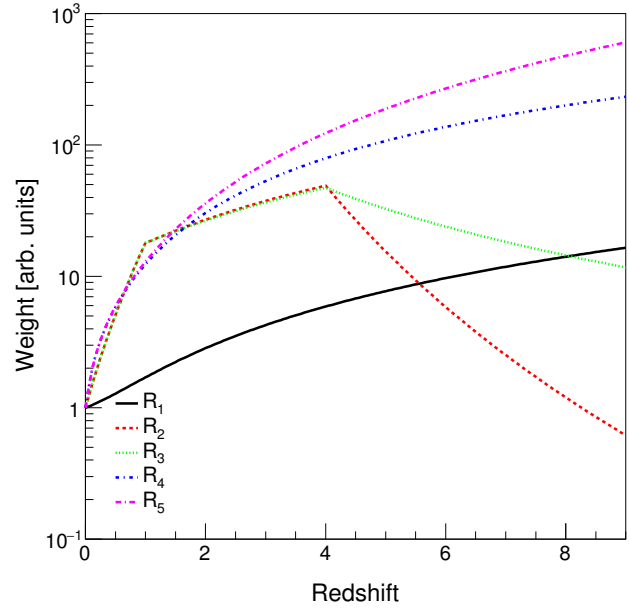}
	\caption{Source evolution with redshift. Each line represents one of the models $R_{n}$, see text for details of the models.}
	\label{fig:redshift}
\end{figure}

Figure~\ref{fig:redshift} shows the ratio of sources as a function of redshift for the five source distributions considered. The source evolution uniformly distributed in a comoving volume is shown only for comparison. It is clear that even astrophysical motivated evolutions are different for redshift larger than two. Charged particles produced in sources farther than redshift equals to one have a negligible probability of reaching Earth, however the GZK photons produced in their propagation could travel farther if \liv is considered.

\begin{figure}[!h]
	\centering
	\includegraphics[width=.48\textwidth]{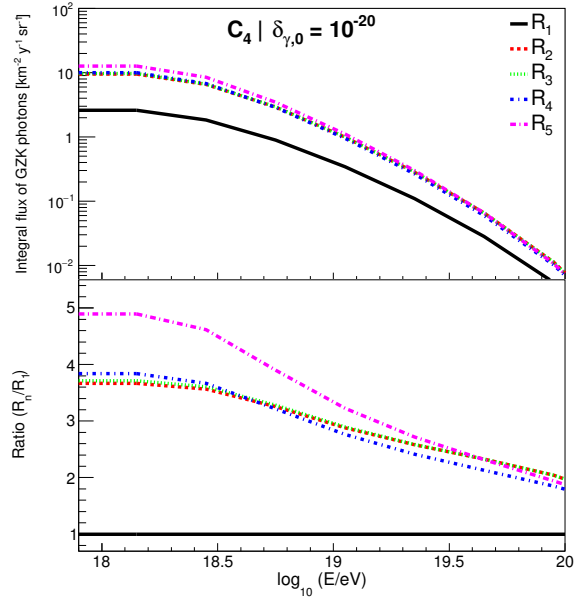}
	\caption{Integral flux of GZK photons as a function of the photon energy for each source evolution model. Each line represents a different model $R_{n}$. All cases are for the source model $C_{4}$ and \liv coefficient $\delta_{\gamma,0} = 10^{-20}$. The top panel shows the integral flux, while the bottom panel show the ratio to the simplest case, $R_{1}$.}
	\label{fig:flux:redshift}
\end{figure}

Figure~\ref{fig:flux:redshift} shows the effect of the source evolution in the prediction of GZK photons including \liv effects. Once more, the use of different \liv coefficients results in a shift up an down in the integral flux for each source evolution model, having negligible changes in each ratio. The differences for each source evolution model are as large as 500\% at $E = 10^{18}$ eV. The capability to restrict \liv effects is proportional to the GZK photon flux generated in each model assumption.

\section{Limits on \liv coefficients}
\label{sec:limits}

The GZK photon flux of the five astrophysical models shown above are considered together with the upper limits on the photon flux imposed by the Pierre Auger Observatory to set limits on the \liv coefficients. The simulations considered sources up to 9500 Mpc ($z \approx 8.88$). The reference results are for model $C_3R_{5}$, as this is the model which best describes current UHECR data. The three orders of \liv ($n =$ 0,1 and 2) are considered for each astrophysical model $C_{i}$. Two limiting cases are also considered: \li and maximum \liv, labeled as $\delta_{\gamma}=0$ and $\delta_{\gamma} \rightarrow - \infty$, respectively. The Lorentz Invariant case (\li) is shown for comparison. The maximum \liv case ($\delta_{\gamma} \rightarrow - \infty$) represents the limit in which the mean free path of the photon-photon interaction goes to infinity at all energies and therefore no interaction happens. These two cases bracket the possible \liv solutions. The UHECR flux reaching Earth was normalized to the flux measured by the Pierre Auger Observatory (\cite{bib:AugerSpectrum}) at $E = 10^{18.75}$~eV which sets the normalization of the GZK photon flux produced in the propagation of these particles.

\begin{figure}[!h]
	\centering
	\includegraphics[width=.7\textwidth]{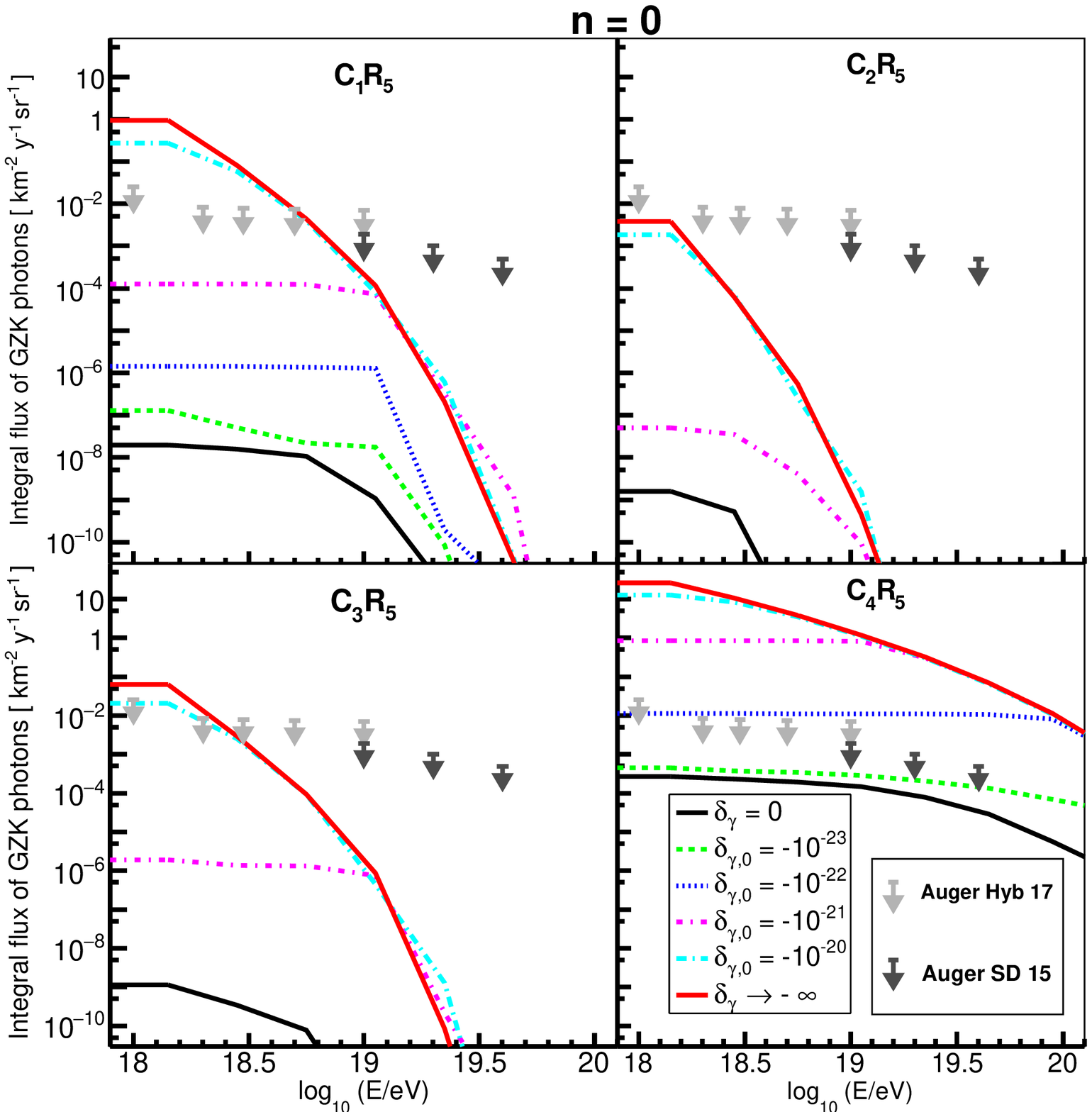}
	\caption{Integral flux of GZK photons as a function of the photon energy considering \liv effects for $n = 0$. The black continuous line represents the \li scenario. The colored lines represent different values for the \liv coefficients. The red line represents the limit LIV case. The arrows represent the upper limits from the Pierre Auger Observatory. Each panel represent a source model, $C_{1}R_{5}$, $C_{2}R_{5}$, $C_{3}R_{5}$, $C_{4}R_{5}$, respectively.}
	\label{fig:Spectran0}
\end{figure}

\begin{figure}[!h]
	\centering
	\includegraphics[width=.7\textwidth]{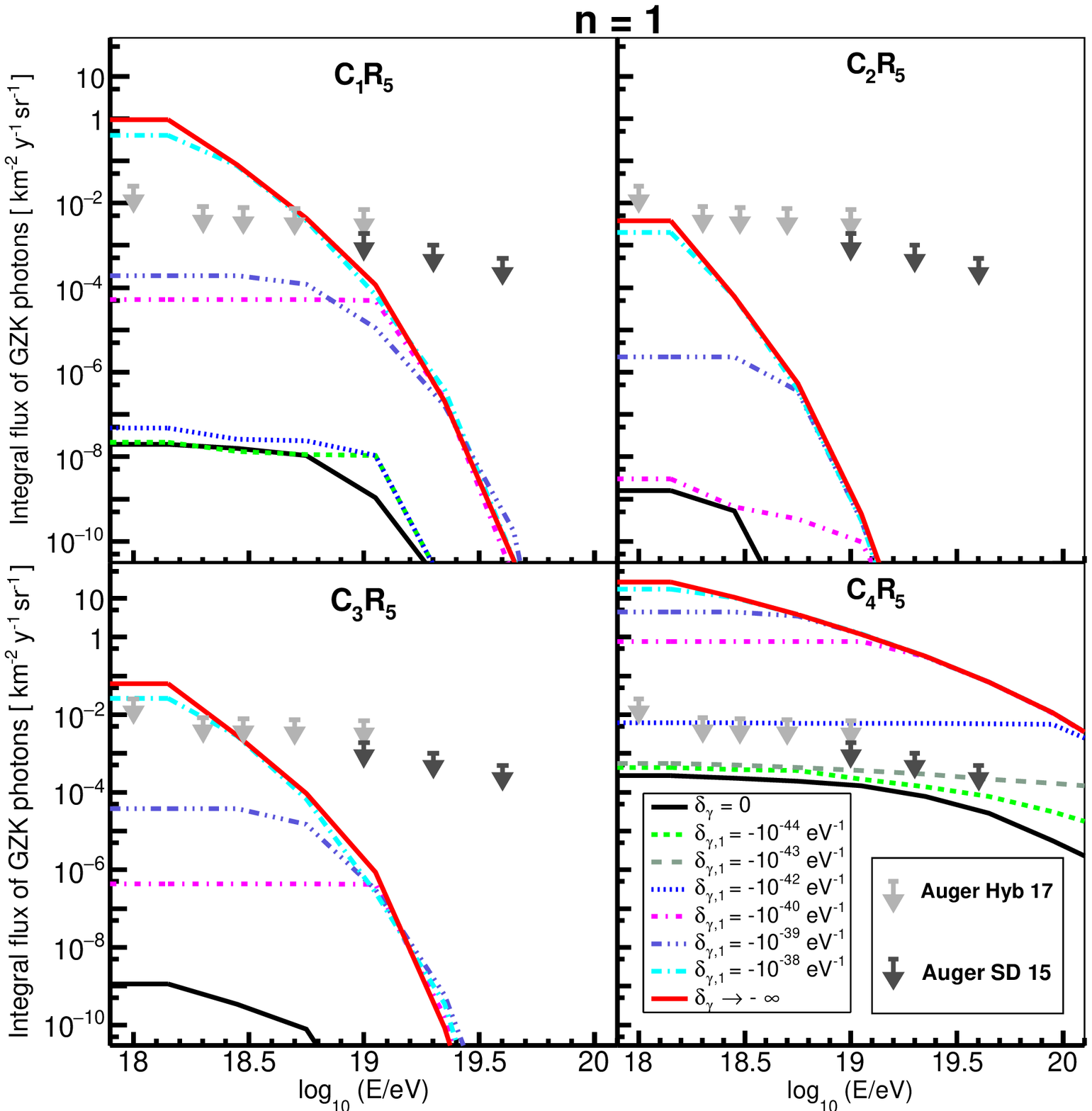}
	\caption{Integral flux of GZK photons as a function of the photon energy considering \liv effects for $n = 1$. The black continuous line represents the \li scenario. The colored lines represent different values for the \liv coefficients. The red line represents the limit LIV case. The arrows represent the upper limits from the Pierre Auger Observatory. Each panel represent a source model, $C_{1}R_{5}$, $C_{2}R_{5}$, $C_{3}R_{5}$, $C_{4}R_{5}$, respectively.}
	\label{fig:Spectran1}
\end{figure}

\begin{figure}[!h]
	\centering
	\includegraphics[width=.7\textwidth]{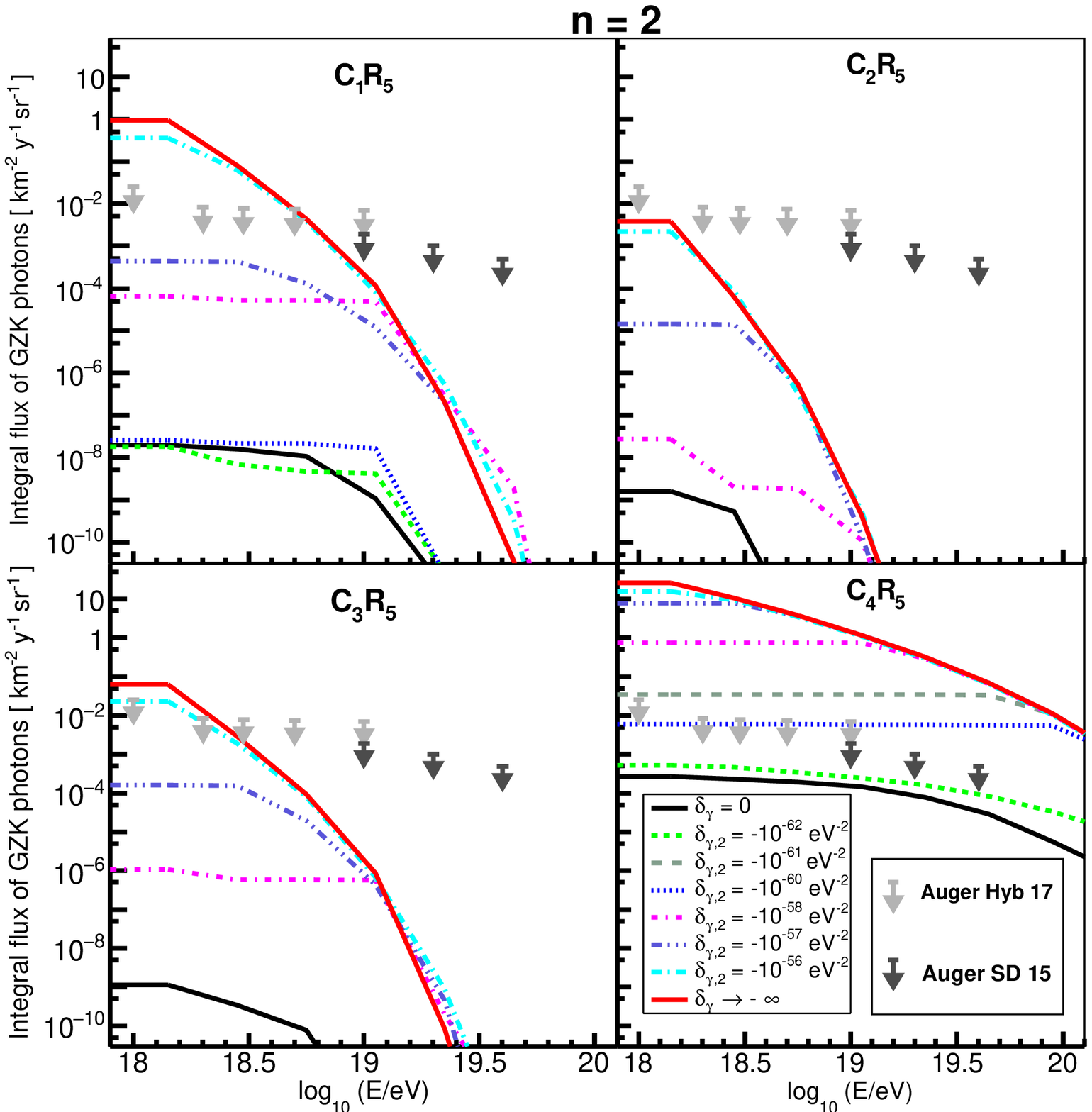}
	\caption{Integral flux of GZK photons as a function of the photon energy considering \liv effects for $n = 2$. The black continuous line represents the \li scenario. The colored lines represent different values for the \liv coefficients. The red line represents the limit LIV case. The arrows represent the upper limits from the Pierre Auger Observatory. Each panel represent a source model, $C_{1}R_{5}$, $C_{2}R_{5}$, $C_{3}R_{5}$, $C_{4}R_{5}$, respectively.}
	\label{fig:Spectran2}
\end{figure}

Figures~\ref{fig:Spectran0}-\ref{fig:Spectran2} show the results of the calculations. For some \liv coefficients, models $C_{1}R_{5}$, $C_{3}R_{5}$ and $C_{4}R_{5}$ produces more GZK photons than the upper limits imposed by Auger, therefore, upper limits on the \liv coefficients can be imposed. Model $C_{2}R_{5}$ produces less GZK photons than the upper limits imposed by Auger even for the extreme scenario $\delta_{\gamma} \rightarrow - \infty$, therefore no limits on the \liv coefficients could be imposed. Table~\ref{table:limits} shows the limits imposed in this work for each source model and \liv order.

Table~\ref{table:otherlimits} shows the limits imposed by other works for the photons sector for comparison. The direct comparison of the results obtained here ($C_3R_5$) is only possible to \cite{Gunter:2007} (first line in table~\ref{table:otherlimits}) because of the similar technique based on GZK photons. The differences between the calculations presented here and the limits imposed in reference~\cite{Gunter:2007} can be explained by: a) the different assumptions considered in the $\gamma \gamma$ interactions with \liv, b) the different astrophysical models used and c) the upper limit on the GZK photon flux used. In reference~\cite{Gunter:2007}, the limits were obtained by calculating the energy in which the interaction of a high energy photon with a background photon at the peak of the CMB, i.e., with energy $\epsilon = 6 \times 10^{-4}$~eV, becomes kinematically forbidden. In this work, a more complete approach was used, where the energy threshold was calculated, the mean free path was obtained by integrating the whole background photon spectrum and the propagation was simulated, obtaining the intensity of the flux of GZK photons. The astrophysical scenario used in reference~\cite{Gunter:2007} was a pure proton composition with energy spectrum normalized by the AGASA measurement (\cite{AGASA}) and index $\Gamma = 2.6$. The source distribution was not specified in the study. However, this astrophysical scenario is ruled out by the $X_{\mathrm{max}}$ measurements from the Pierre Auger Observatory (\cite{bib:xmax1,bib:xmax2}). In the calculations presented here, the \liv limits were updated using astrophysical scenarios compatible to the Auger $X_{\mathrm{max}}$ data. Finally, in this paper new GZK photons limits published by Auger are used. The \liv limits presented here are, therefore, more realistic and up to date.

The other values in table~\ref{table:otherlimits} are shown for completeness. The second and third entries are based on energy dependent arrival time of TeV photons: a) a PKS 2155-304 flare measured with H.E.S.S. (\cite{bib:hesspks}) and b) GRB 090510 measured with Fermi-LAT (\cite{bib:fermigrb}). Entry H.E.S.S. - Mrk 501 (2017) (\cite{bib:hessmkr}) in table~\ref{table:otherlimits} is based on the kinematics of the interactions of photons from Mrk 501 with the background. All the studies shown in table~\ref{table:otherlimits} assumes \liv only in the photon sector. However, the systematics of the measurements and the energy of photons (TeV photons versus EeV photons) are very different and a direct comparison between the GZK photon calculations shown here and the time of arrival of TeV photon is not straight-forward.

\begin{table}
\centering
\begin{tabular}{ c | c c c c c }
\hline
\hline
Model & $\delta_{\gamma,0}^{limit}$ & & $\delta_{\gamma,1}^{limit} [\mathrm{eV^{-1}}]$ & & $\delta_{\gamma,2}^{limit} [\mathrm{eV^{-2}}]$ \\
\hline
$C_{1}R_{5}$ & $\sim -10^{-20}$ & & $\sim -10^{-38}$ & & $\sim -10^{-56}$ \\
$C_{2}R_{5}$ & - & & - & & - \\
$C_{3}R_{5}$ & $\sim -10^{-20}$ & & $\sim -10^{-38}$ & & $\sim -10^{-56}$ \\
$C_{4}R_{5}$ & $\sim -10^{-22}$ & & $\sim -10^{-42}$ & & $\sim -10^{-60}$ \\
\hline
\hline
\end{tabular}
  \caption{Limits on the \liv coefficients imposed by this work for each source model and \liv order ($n$). Model $C_3R_5$ is pointed as the reference values of this paper because it is able to describe the current UHECR data.}
  \label{table:limits}
\end{table}

\begin{table}
\centering
\begin{tabular}{ c | c c c c c }
\hline
\hline
Model & $\delta_{\gamma,0}^{limit}$ & & $\delta_{\gamma,1}^{limit} [\mathrm{eV^{-1}}]$ & & $\delta_{\gamma,2}^{limit} [\mathrm{eV^{-2}}]$ \\
\hline
Galaverni \& Sigl (2008) & - & & $-1.97 \times 10^{-43}$ & & $-1.61 \times 10^{-63}$ \\ \hline
H.E.S.S. - PKS 2155-304 (2011) & - & & $-4.76 \times 10^{-28}$ & & $-2.44 \times 10^{-40}$ \\
Fermi - GRB 090510 (2013) & - & & $-1.08 \times 10^{-29}$ & & $-5.92 \times 10^{-41}$ \\
H.E.S.S. - Mrk 501 (2017) & - & & $-9.62 \times 10^{-29}$ & & $-4.53 \times 10^{-42}$ \\
\hline
\hline
\end{tabular}
  \caption{Limits on the \liv coefficients imposed by other works based on gamma-ray propagation. First line shows a previous result which can be directly compared to the calculations presented here in table~\ref{table:limits}. Last three lines are shown for completeness. These limits are based on gamma-ray arrival time and are not directly comparable to the ones in table~\ref{table:limits}.}
  \label{table:otherlimits}
\end{table}

\section{Conclusions}
\label{sec:conclusions}

In this paper, the effect of possible \liv in the propagation of photons in the Universe is studied. The interaction of a high energy photon traveling in the photon background was solved under \liv in the photon sector hypothesis. The mean free path of the $\gamma \gamma_{CB} \rightarrow e^{+} e^{-}$ interaction was calculated considering \liv effects. Moderate \liv coefficients introduce a significant change in the mean free path of the interaction as shown in section~\ref{sec:horizon} and figures~\ref{fig:LIVpathn0},~\ref{fig:LIVpathn1} and~\ref{fig:LIVpathn2}. The corresponding \liv photon horizon was calculated as shown in figure~\ref{fig:LIVhorizon}.

The dependence of the integral flux of GZK photons on the model for the sources of UHECRs is discussed in section~\ref{sec:gzk:photons} and shown in figures~\ref{fig:source:models} and \ref{fig:flux:redshift}. The flux changes several orders of magnitude for different injection spectra models. A difference of about 500\% is also found for different source evolution models. Previous \liv limits were calculated using GZK photons generated by source models currently excluded by the data (\cite{Gunter:2007}). The calculations presented here shows \liv limits based on source models compatible with current UHECR data. In particular, model $C_3R_5$ was shown to describe the energy spectrum, composition and arrival direction of UHECR~(\cite{bib:C2}) and therefore is chosen as our reference result.

The calculated GZK photon fluxes were compared to most updated upper limits from the Pierre Auger Observatory and are shown in figures~\ref{fig:Spectran0}-\ref{fig:Spectran2}. For some of the models, it was possible to impose limits on the \liv coefficients, as shown in table~\ref{table:limits}. It is important to note that the \liv limits shown in table~\ref{table:limits} were derived from astrophysical models of UHECR compatible to the most updated data. The limits presented here are several order of magnitudes more restrictive than previous calculations based on the arrival time of TeV photons (\cite{bib:hesspks,bib:fermigrb}), however, the comparison is not straight-forward due to different systematics of the measurements and energy of the photons.

\acknowledgments

RGL is supported by FAPESP (2014/26816-0, 2016/24943-0). HMH acknowledges IFSC/USP for their hospitality during the developments of this work, Abdel P\'erez Lorenzana for enlightening discussions and the support from Conacyt Mexico under grant 237004 and the Brazilian agency FAPESP (2017/03680-3). VdS thanks the Brazilian population support via FAPESP (2015/15897-1) and CNPq. This work has partially made use of the computing facilities of the Laboratory of Astroinformatics (IAG/USP, NAT/Unicsul), whose purchase was made possible by the Brazilian agency FAPESP (2009/54006-4) and the INCT-A.The authors acknowledge the National Laboratory for Scientific Computing (LNCC/MCTI, Brazil) for providing HPC resources of the SDumont supercomputer, which have contributed to the research results reported within this paper (\url{http://sdumont.lncc.br}).

\bibliographystyle{aasjournal}
\bibliography{paper}

\appendix
\section{Description of the \liv model}
\label{sec:liv:model}

Equation~\ref{eq:dispersion} leads to unconventional solutions of the energy threshold in particle production processes of the type $AB\rightarrow CD$. In this paper, the $\gamma\gamma_{CB}\rightarrow e^+ e^-$ interaction is considered. From now on, the symbol $\gamma$ refers to a high energy gamma ray with energy $E_\gamma = [10^{9},10^{22}]$ eV that propagates in the Universe and interacts with the cosmic background (CB) photons,  $\gamma_{CB}$, with energy $\epsilon = [10^{-11},10]$ eV.

Considering \liv in the photon sector, the specific dispersion relations can be written:
\begin{equation}
\begin{aligned}
E_{\gamma}^2 - p^2_{\gamma} =  \delta_{\gamma,n} \; E_{\gamma}^{n+2}, \\
\epsilon^2 - p^2_{\gamma_{CB}} =  \delta_{\gamma,n} \; \epsilon^{n+2},
\end{aligned}
\label{eq:s:gamma}
\end{equation}
where $\delta_{\gamma,n}$ is the $n$-order \liv coefficient in the photon sector and therefore taken to be the same in both dispersion relations. The standard \li dispersion relation for the electron-positron pair follows: $E_{e^\pm}^2 - p^2_{e^\pm} = m_e^2.$

Taking into account the inelasticity ($K$) of the process ($E_{e^-} = K E_\gamma$) and imposing energy-momentum conservation in the interaction, the following expression for a head-on collision with collinear final momenta can be written to leading order in $\delta_{\gamma,n}$
\begin{equation}
\begin{aligned}
& 4\epsilon E_{\gamma}  - m_e^2 \left( \frac{1}{K(1-K)} - \frac{m_e^2}{2K(1-K)(E_{\gamma}+\epsilon)^2} \right)
& = - \delta_{\gamma,n}E_\gamma^{n+2}\left[ 1 + \frac{\epsilon^{n+2}}{E^{n+2}_{\gamma}}
-\frac{\epsilon}{E_{\gamma}} \left( 1+ \frac{\epsilon^n}{E_{\gamma}^n}\right)  \right] \; .
\end{aligned}
\label{eq:e:conservation}
\end{equation}

In the ultra relativistic limit $E_{\gamma} \gg m_e$ and $E_{\gamma} \gg \epsilon$, this equation reduces to
\begin{equation}
 \delta_{\gamma,n} E_{\gamma}^{n+2} + 4E_{\gamma}\epsilon - m_e^2  \frac{1}{K(1-K)} = 0 .
\label{eq:e:cons:lim}
\end{equation}

Equation~\ref{eq:e:cons:lim} implies two scenarios: I) $\delta_{\gamma,n} > 0$ the photo production threshold energy is shifted to lower energies and II) $\delta_{\gamma,n} < 0$ the threshold takes place at higher energies than that expected in a \li regime, except for scenarios below a critical value for delta where the photo production process is forbidden. Notice that, if $\delta_{\gamma,n} = 0$ in equation~\ref{eq:e:cons:lim} the \li regime is recovered. In the \li regime, it is possible to define $E^\li_\gamma = \frac{m_e^2}{4 \epsilon K(1-K)}$. The math can be simplified by the introduction of the dimensionless variables
\begin{equation}
x_{\gamma}=\frac{E_{\gamma}}{E_{\gamma}^\li},
\end{equation}
and
\begin{equation}
\Lambda_{\gamma,n} = \frac{ E_{\gamma}^{LI~(n+1)}}{4\epsilon} \delta_{\gamma,n}.
\end{equation}
Then, equation~\ref{eq:e:cons:lim} takes the form
\begin{equation}
\Lambda_{\gamma,n}x_\gamma^{n+2}+  x_\gamma -1 = 0.
\label{eq:pol}
\end{equation}
Studying the values of $\delta_{\gamma,n}$ for which equation~\ref{eq:pol} has a solution, one can set the extreme allowed \liv coefficient (\cite{Gunter:2008,HMartinez:2016}). The limit \liv coefficient ($\delta_{\gamma,n}^{lim}$) for which the interaction is kinematically allowed for a given $E_\gamma$ and $\epsilon$ is given by:
\begin{equation}
\delta_{\gamma,n}^{lim} =- 4 \frac{\epsilon}{E_\gamma^{LI~(n+1)}}\frac{(n+1)^{n+1}}{(n+2)^{n+2}}.
\label{eq:delta:max}
\end{equation}

Equation~\ref{eq:pol} has real solutions for $x_{\gamma}$ only if $\delta_{\gamma,n} > \delta_{\gamma,n}^{lim}$. Therefore, under the \liv model considered here, if $\delta_{\gamma,n} < \delta_{\gamma,n}^{lim}$, high energy photons would not interact with background photons of energy $\epsilon$.

\begin{figure}[!ht]
    \centering
	    \includegraphics[width=.6\textwidth]{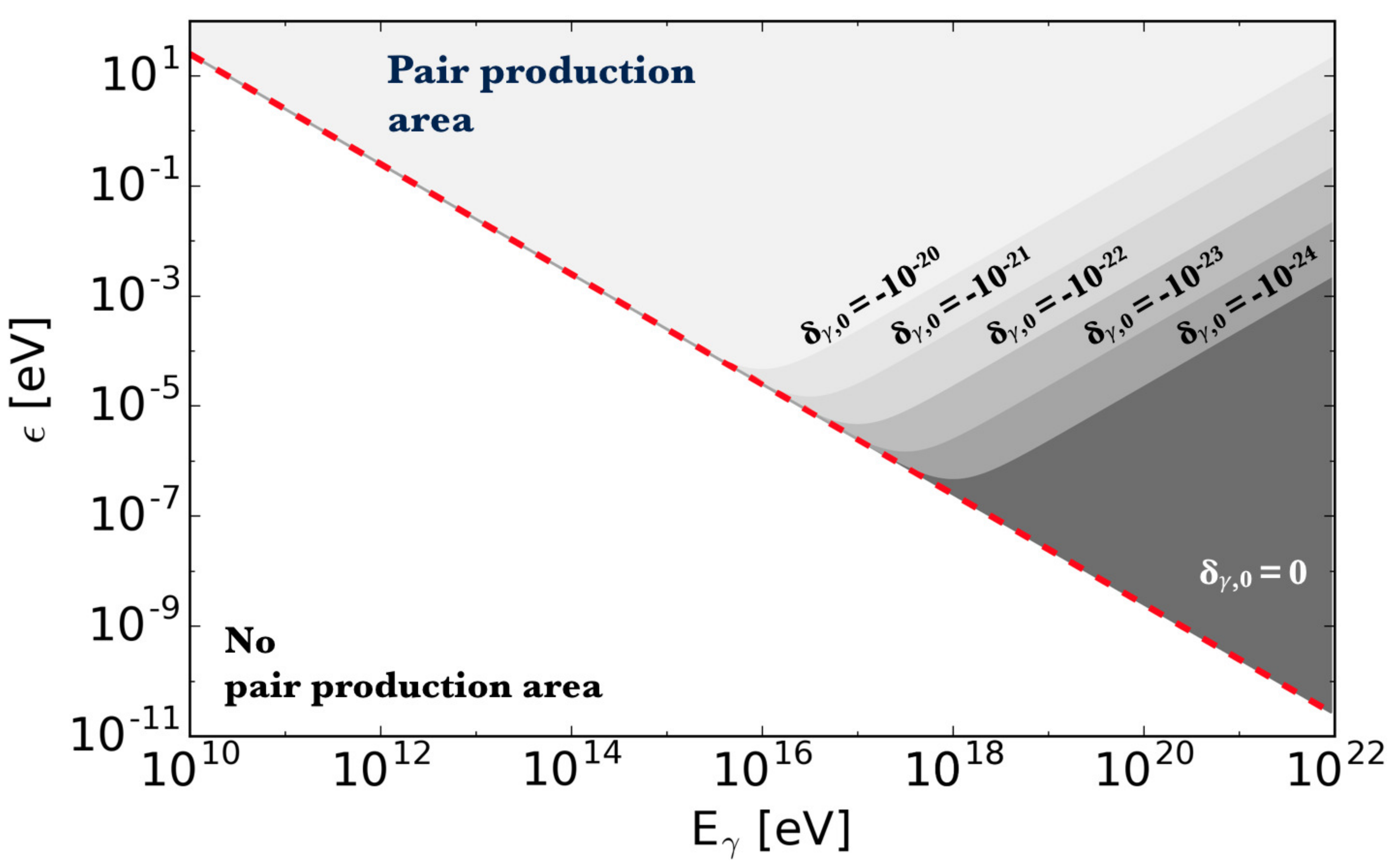}
        \caption{Allowed regions for the pair production in the $\gamma\gamma_{CB}\rightarrow e^+ e^-$ interaction considering \liv effects. The high energy photon ($E_\gamma$) and background photon ($\epsilon$) parameter space is shown divided in gray regions for each value of $\delta_{\gamma,0}$. The gray areas are cumulative from darker to lighter gray. The red dashed line is a reference for $\delta_{\gamma,0} = 0$ (\li case).}
        \label{fig:pair:production}
\end{figure}

For a given $E_\gamma$ and $\delta_{\gamma,n}$ the threshold background photon energy ($\epsilon_{th}^\liv$) including \liv effects is:
\begin{equation}
\epsilon_{th}^\liv  =  \frac{m_e^2}{4E_{\gamma}K(1-K)} -  \frac{\delta_{\gamma,n} E_{\gamma}^{n+1}}{4}.
\label{eq:e:cons:th}
\end{equation}
The superscript \liv is used for emphasis. In the paper, $\epsilon_{th}^\liv$ as given by equation~\ref{eq:e:cons:th} will be used for the calculations of the mean free path of the $\gamma\gamma_{CB}\rightarrow e^+ e^-$ interaction. Figure~\ref{fig:pair:production} shows the allowed parameter space of $E_\gamma$ and $\epsilon$ for different values of $\delta_{\gamma,0}$. The gray areas are cumulative from darker to lighter gray.

\newpage


\end{document}